\documentclass[letterpaper]{article} 
\usepackage{aaai25}  
\usepackage{times}  
\usepackage{helvet}  
\usepackage{courier}  
\usepackage[hyphens]{url}  
\usepackage{graphicx} 
\usepackage{natbib}  
\usepackage{caption} 
\usepackage{algorithm}
\usepackage{algorithmic}
\usepackage{comment}

\usepackage{newfloat}
\usepackage{listings}
\DeclareCaptionStyle{ruled}{labelfont=normalfont,labelsep=colon,strut=off} 
\floatstyle{ruled}
\newfloat{listing}{tb}{lst}{}
\floatname{listing}{Listing}
\title{Second-order Theory of Mind for Human Teachers and Robot Learners.}

\author {
    Patrick Callaghan, 
    Reid Simmons, 
    Henny Admoni 
}
\affiliations {
    Carnegie Mellon University\\
    callaghan@cmu.edu  
}

\begin{document}

\maketitle

\section{Introduction}

Confusing or otherwise unhelpful learner feedback creates or perpetuates erroneous beliefs that the teacher and learner have of each other, thereby increasing the cognitive burden placed upon the human teacher. For example, the robot's feedback might cause the human to misunderstand what the learner knows about the learning objective or how the learner learns. At the same time---and in addition to the learning objective---the learner might misunderstand how the teacher perceives the learner's task knowledge and learning processes. To ease the teaching burden, the learner should provide feedback that accounts for these misunderstandings and elicits efficient teaching from the human.

One way to account for these erroneous beliefs and thereby improve a human's teaching efficacy is to leverage Theory of Mind. Theory of Mind (ToM)---the ability to infer another's motives and beliefs by observing their actions---is often used in AI approaches today~\cite{premackDoesChimpanzeeHave, yuanSituBidirectionalHumanrobot2022, grayManipulatingMentalStates2014, celikokInteractiveAITheory2019, oguntolaDeepInterpretableModels2021}. Less explored, however, is \emph{Second-order} Theory of Mind (ToM-2) which includes an awareness that other agents also have a ToM~\cite{astingtonTheoryMindEpistemological2002, paperaDevelopmentSecondorderTheory2019}. With this additional awareness, a learner could model and account for its teacher's beliefs of the learner when selecting feedback during a teaching session.

This work endows an AI learner with a ToM-2 that models perceived rationality as a source for the erroneous beliefs a teacher and learner may have of one another. It also explores how a learner can ease the teaching burden and improve teacher efficacy if it selects feedback which accounts for its model of the teacher's beliefs about the learner \emph{and} its learning objective.

\vspace{-1mm}
\section*{Proof of Concept Problem Domain}

\begin{figure}[h!]
    \centering
    \includegraphics[scale=0.35]{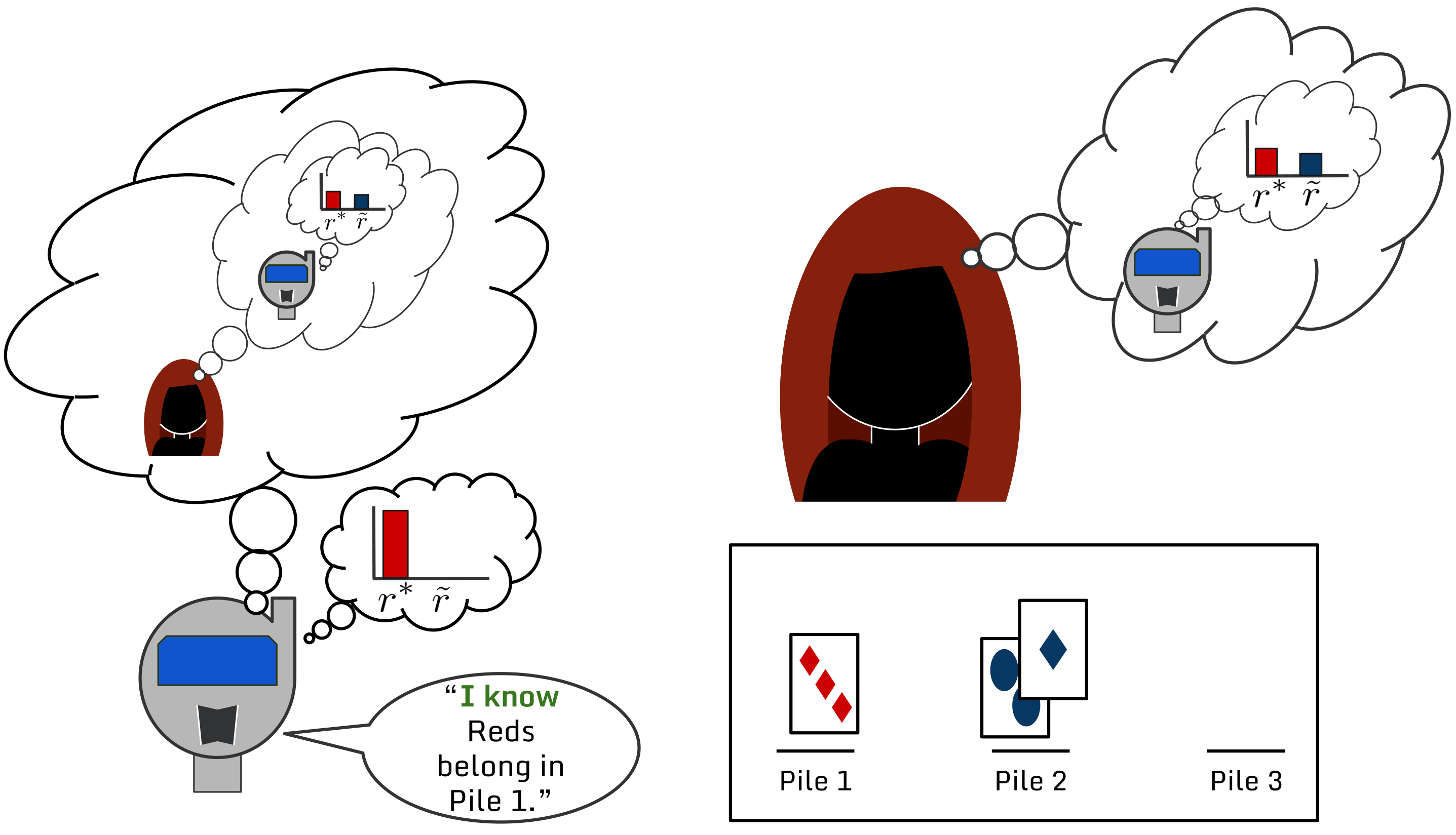}
    \caption{A human teaches a robot the $[r]$ule which dictates how cards are categorized (e.g., according to color). Here, the teacher misunderstands if the robot knows the correct rule $r^*$. The robot's Second-order Theory of Mind enables it to model this misunderstanding and provide feedback using Confidence Expressions (green text).}
    \label{fig:set}
\end{figure}

Consider a turn-based card game played between a human teacher and robot learner in which a ``rule'' governs how multi-featured cards are sorted into three piles (Figure~\ref{fig:set}). In a single round, the teacher plays one such card into a pile according to the rule, and the robot responds with an utterance (i.e., ``feedback'') pertaining to one of the features of the rule. The robot's goal is to identify the rule which distinguishes the piles.

Let's say the teacher chooses the rule: ``Reds belong in Pile 1. Blues belong in Pile 2. Greens belong in Pile 3.'' In the first round of the game, the teacher places the ``Three Red Diamonds'' card on Pile 1. What should the learner infer from this move, and what feedback will convey the learner's belief and prompt the teacher to play an informative next card? Prior work endows the learner with policies for selecting feedback that optimize volume removal and information gain~\cite{fitzgeraldINQUIREINteractiveQuerying2022, sadighActivePreferenceBasedLearning2017, palanLearningRewardFunctions2019, holladayActiveComparisonBaseda, biyikAskingEasyQuestions2019}. Unfortunately, such optimizations can yield feedback that causes the teacher to misunderstand what the learner knows about the task or how the learner incorporates information into its reasoning processes. In the context of the game, such optimizations can yield redundant feedback that reflects stunted learning (e.g., the robot selects the same utterances at turn $t$ and turn $t+1$), seemingly-irrelevant feedback that reflects incorrect learning (e.g., the robot selects an utterance about Pile 3 when the teacher has focused on Pile 1), or feedback which otherwise reflects the robot's erroneous beliefs. Crucially, however, the learner may be closer to the truth than the human believes, yet because it optimized its feedback for the learning objective and didn't consider the teacher's beliefs of the learner, the feedback may compel the teacher to re-teach or continue teaching concepts which the robot already mastered. A learner endowed with a ToM-2 could model what the teacher believes of the learner and account for those beliefs when selecting feedback to mitigate such misunderstandings (see Figure~\ref{fig:set}).

\vspace{-2mm}
\section{Methods}
To undertake the proposed solution, this work leverages the Interactive Partially Observable Markov Decision Process (I-POMDP) as a framework for a robot learner's ToM-2~\cite{gmytrasiewiczFrameworkSequentialPlanning2005}. The I-POMDP augments the POMDP's notion of state to an \emph{interactive state}, which represents both the states of the environment \emph{and} models of the other agents within it. These models can be I-POMDPs which themselves represent environment states and models of other agents, and it is this configuration that enables the I-POMDP to represent a Second-order Theory of Mind.

With this framework, the learner can model the teacher's beliefs about the learner's task knowledge and learning processes through the components which comprise an I-POMDP---namely, an agent's observation function, reward function, transition function, and optimization criterion. Of these elements, this work posits that erroneous beliefs primarily stem from the teacher's observation function. More specifically, a human teacher might observe and interpret a learner's feedback in a perfectly-rational manner, a perfectly-irrational manner, or somewhere in between. By modeling these different possibilities for how the teacher perceives the learner's feedback, the learner can come to a better understanding of how the the teacher will interpret its feedback.

At the same time, the teacher may believe the learner observes and interprets the teacher's actions in an irrational fashion. Indeed, prior work suggests this perceived-irrationality can manifest as the teacher outright ignoring the learner's feedback and playing cards in a systematic fashion~\cite{koppolInteractiveMachineLearning}. By modeling the teacher's belief of the learner's observation function, the learner can identify if the teacher believes the robot is irrational and choose feedback most likely to lead the teacher to understand the robot is, in fact, a rational learner. In turn, the teacher could become more efficient in how they play cards.

To enable the learner to recognize these sources of irrationality, this work augments the I-POMDP with a discrete set of learnable observation functions, each of which is a noisily-rational model whose rationality is inversely-proportional to a temperature parameter $\beta$. By observing the teacher's card plays, the learner will be able to learn the teacher's degree of rationality as it also learns the rule governing the teacher's actions.

Modeling these sources of irrationality is only helpful to the teacher if the learner accounts for them when generating feedback. As such, this work incorporates \emph{Confidence Expressions} (CEs) into the learner's feedback to convey the strength of the learner's stated beliefs. Without CEs, feedback can express confidence far greater than the learner's actual confidence about the features it addresses, and this ambiguous communication can lead the teacher to misunderstand the learner's task knowledge and learning processes. For example, if the learner says, ``Reds belong in Pile 1,'' the teacher might perceive the learner as 100\% certain of its statement. It's possible, however, that the learner is still unsure if Reds or Diamonds categorize that Pile.  

CEs are meant to convey the learner's certainty about the stated features. More specifically, the learner prepends one of three CEs (``I know,'' ``I think,'' or ``I'm unsure if'') to its feature expression (e.g., ``Reds belong in Pile 1.''), selecting the one most reflective of its confidence in the stated feature. By incorporating CEs into its feedback, the learner is able to convey its level of certainty over its task knowledge with the intention of correcting and preventing teacher misunderstandings. If we again consider Figure~\ref{fig:set}, perhaps the teacher believes the learner needs to see the complete set of Red cards in Pile 1 to completely understand that aspect of the rule. This strategy, however, is redundant and elicits unnecessary time investment on the part of the teacher. To mitigate this extra effort, the robot could say, ``I know Reds belong in Pile 1,'' when certain of that feature. The robot can additionally express its uncertainty over other features by stating, for example, ``I'm unsure if Greens belong in Pile 3.''

\vspace{-3.5mm}
\section{Evaluations}
This work will evaluate the utility of endowing a robot learner with a ToM-2 through both simulated and real-world interactions between teacher and learner. The evaluations will investigate the benefits of enabling a learner to (1) identify its teacher's sources of irrationality and (2) utilize Confidence Expressions when providing feedback during the teaching session. The turn-based card game will be the domain of study.
\paragraph{Simulation Experiments} The first set of experiments will comprise interactions between a learner and a simulated teacher to investigate the learner's ability to identify the teacher's rationality and the teacher's perceived learner rationality. They will also investigate if this knowledge enables the learner to elicit greater teaching efficacy from its teacher. Each trial will initialize a teacher with a noisily-rational observation function, and the learner's inferred model will be compared against the ground truth teacher model for the span of the teaching session. Additionally, experiments will evaluate how CEs benefit teacher efficacy, the learner's ability to identify a teacher's rationality, and if their use can guide a teacher to better understanding of the learner's rationality.
\paragraph{User Study} The next set of experiments will be undertaken through a user study in which human participants teach a robot learner rules of varied complexities. As in the simulated interactions, the user study will quantify these benefits through the number of rounds it takes the robot to learn the rule in each experimental condition, as well as the number of times the teacher incorrectly believes the learner understands the rule. Additionally, subjective metrics (e.g., the NASA TLX) will be used to evaluate the cognitive burden imposed by each of the conditions~\cite{hartDevelopmentNASATLXTask1988}.

\bibliography{references}

\end{document}